\documentclass[useAMS,usenatbib]{mn2e}

\usepackage{graphicx}
\def\etal{et~al.}
\def\kms{\,km\,s$^{-1}$}
\def\kmsd{\,km\,s$^{-1}$\,d$^{-1}$}

\def\M{M$_{\odot}$}

\def\CII{C\,{\sc ii}}
\def\OI{O\,{\sc i}}

\def\SII{S\,{\sc ii}}
\def\SIII{S\,{\sc iii}}

\def\SiII{Si\,{\sc ii}}
\def\SiIII{Si\,{\sc iii}}
\def\MgII{Mg\,{\sc ii}}
\def\CaII{Ca\,{\sc ii}}
\def\FeII{Fe\,{\sc ii}}
\def\FeIII{Fe\,{\sc iii}}
\def\CoII{Co\,{\sc ii}}
\def\CoIII{Co\,{\sc iii}}
\def\NiII{Ni\,{\sc ii}}
\def\TiII{Ti\,{\sc ii}}
\def\CrII{Cr\,{\sc ii}}
\def\CrIII{Cr\,{\sc iii}}

\def\Ang{\,\AA}
\def\Nifs{$^{56}$Ni}
\def\Nife{$^{58}$Ni}
\def\Cofs{$^{56}$Co}
\def\Fefs{$^{56}$Fe}
\def\Feff{$^{54}$Fe}
\def\Fefe{$^{58}$Fe}

\def\2bo{SN~2002bo}
\def\ergs{\,(erg\,s$^{-1}$)}
\title[Abundance Stratification in Type Ia Supernovae:
{\sc i}. The case of SN~2002bo]
{Abundance Stratification in Type Ia Supernovae:\\
{\sc i}. The Case of SN~2002bo}
\author[M. Stehle et al.]
{M. Stehle$^{1,2}$, P. A. Mazzali$^{3,1}$, S. Benetti$^{4}$,
W. Hillebrandt$^{1}$\\
$^{1}$Max-Planck-Institut f\"ur Astrophysik, Karl-Schwarzschild-Str. 1,
D-85741 Garching bei M\"unchen, Germany\\
$^{2}$Universit\"ats-Sternwarte M\"unchen, Scheinerstr. 1,
D-81679 M\"unchen, Germany\\
$^{3}$Osservatorio Astronomico di Trieste, Via Tiepolo 11,
I-34131 Trieste, Italy\\
$^{4}$INAF, Osservatorio Astronomico di Padova, Vicolo
dell'Osservatorio 5, I-35122 Padova, Italy}
\begin{document}
\date{Accepted ... Received ...;}
\pagerange{\pageref{firstpage}--\pageref{lastpage}} \pubyear{2004}
\maketitle
\label{firstpage}
\begin{abstract}

The abundance stratification in the ejecta of the normal Type Ia
Supernova 2002bo is derived fitting a series of spectra obtained at
close time intervals. A Montecarlo code, modified to include
abundance stratification, is used to compute synthetic spectra at 13
epochs in the photospheric phase, starting 13 days before $B$
maximum. A description of the abundance distribution above 7600\kms\
is thus obtained. Abundances in deeper layers, down to zero
velocity, are derived from models of two nebular-phase spectra.
Elements synthesised in different stages of burning are
significantly, but not completely mixed in the ejecta. A total
\Nifs\ mass of 0.52\,\M\ is derived. Evidence for intermediate-mass
elements at high velocities ($\ga 18,000$\kms) is found, most
clearly in \SiII\ 6355\Ang, \CaII\ H\&K, and in the \CaII\ IR
triplet. Carbon lines are not seen at any velocity, with possible
implications on the progenitor/explosion scenario. A synthetic
bolometric light curve computed using the inferred abundance
distribution is in very good agreement with the observed one,
providing an independent check. In particular, the fast rise of the
light curve is reproduced very well. This is due to outward mixing
of \Nifs, which is clearly well determined by the spectroscopic
modelling.

\end{abstract}
%
%
\section{Introduction} \label{intro}

It is widely agreed that Type Ia Supernovae (SNe~Ia) result from the
thermonuclear explosion of Carbon-Oxygen white dwarfs (WD) in binary
systems. Two possible progenitor configurations are usually
considered. In the single degenerate scenario a WD accretes mass
from a Roche lobe-filling Red Giant companion. When the WD reaches a
mass close to the Chandrasekhar limit ($\approx$ 1.38\,\M), carbon
burning is triggered by compressional heating near the centre. After
a few thousand years of quiet burning \citep{ib84,we84,w04}, a
thermonuclear runaway occurs which disrupts the star. In the double
degenerate scenario, two low-mass WDs in a close binary system with
total mass exceeding the Chandrasekhar mass lose angular momentum
via the emission of gravitational waves, which ultimately leads to
the merging of the two stars and to a thermonuclear runaway
\citep{wh73,n82,h04}. Potential progenitor systems for both channels
have been detected, but their numbers are too low to explain the
frequency of occurrence of SNe~Ia \citep{ca99}.

Although the apparent predictability and homogeneity of SNe~Ia,
together with their brightness, has motivated their use as
standardisable cosmological candles \citep{pe99,ri98}, much remains
unknown about their properties, as regards both global physical
processes and the peculiarities of individual objects.

For example, the details of the explosion process are still unclear.
Ignition is supposed to start near the centre. A subsonic
(deflagration) wave, often called a ``flame'', travels outwards,
burning part of the WD to nuclear statistical equilibrium (NSE). The
subsonic speed of the deflagration front prevents the WD from being
burned to NSE entirely. Partial burning of C and O results in the
production of intermediate mass elements (IME): Si, S, Mg and Ca
dominate the early spectra of SNe~Ia. The prompt detonation
mechanism \citep{ar69}, on the other hand, is inconsistent with the
spectra, as it fails to produce sufficient amounts of IME. There is
still no agreement as to whether the explosion continues as a
deflagration, becoming strongly turbulent \citep[Nomoto, Thielemann
\& Yokoi 1984;][]{wo84,ni95,re02} or it turns to a supersonic
detonation. Recent 3D deflagration models leave significant amounts
of unburned material both in the outer and in the inner parts of the
ejecta \citep{ga03,re99,t04}, and produce too little radioactive
\Nifs. This could affect the light curve of SNe~Ia, which is powered
by \Nifs\ through the energy released in the decay chain to stable
\Fefs\ via \Cofs. Presently, delayed detonation models
\citep{ho96,i99} describe better the abundances of Fe-group elements
and IME at high velocities, as in the detonation phase burning is
boosted in the outer regions of the envelope. These models also have
the flexibility to produce different amounts of \Nifs\ and IME, but
this depends on the ad-hoc assumption of the time and position of
occurrence of the Deflagration to Detonation Transition
\citep[DDT,][]{k91,ww94}.

Another area that needs clarification is the final distribution of
the elements in the ejecta.  Light curve studies typically make use
of the direct results of the explosion models, where the abundances
are stratified, but spectral analysis has favoured mixing in the
ejecta, at least above some velocity \citep{br85}. Mixing was
favoured over stratification in an LTE study \citep{har91}, but no
further investigations have been performed, although the results
could be very useful to discriminate between different scenarios for
the explosion.

The European Research and Training Network (RTN) "The Physics of
Type Ia Supernova Explosions"  was set up to study SNe~Ia by means
of very good time series of multi-wavelength observations and
detailed models of a sample of nearby objects. \2bo\ was its first
target \citep{be04}. In this paper we model the spectra and light
curve of SN~2002bo, and attempt to extract information about the
abundance stratification in the SN ejecta.

In Section~\ref{obs} we summarise the basic properties of \2bo. The
method of the analysis is discussed in Section~\ref{model}. Models
of the photospheric epoch spectra are presented in
Section~\ref{analysis}, while Section~\ref{neb} focuses on models of
the nebular spectra. The derived abundance distribution is presented
and discussed in Section~\ref{at}. In Section~\ref{lc} a synthetic
light curve computed on the basis of the derived abundance
distribution is presented and discussed. In Section~\ref{vel} we
discuss the behaviour of the line velocities, and in Section~9 the
results are summarised.

%
\section{Observations}\label{obs}

\2bo\ was discovered independently by Cacella and Hirose
\citep{ca02} on UT March 9.08 and 9.505, 2002. It is located in
NGC~3190 at a distance modulus 
$\mu = 31.67$ \citep[$H_0 = 65$\kms$\,{\rm Mpc}^{-1}$]{be04}. 
\citet{be04} present photometry
and spectroscopy covering from $\sim 13$~days before to 368\,days
after the estimated epoch of $B$~maximum.  They quote a $B$-band
risetime of $17.9\,\pm\,0.5$~d, a decline rate $\Delta m_{15}(B) =
1.13\,\pm\,0.05$, a reddening $E(B-V)_{obs} = 0.43\,\pm\,0.10$ and a
reddening-corrected $M_B = -19.41\,\pm\,0.42$, making \2bo a normal SN~Ia
\citep{br93}. However, comparison with other SNe~Ia reveals
peculiarities, e.g. with respect to line velocities.


\section{Method of Analysis}\label{model}

In order to describe the early-time evolution of a SN spectrum (and
in particular of a SN~Ia) it is in principle necessary to solve the
time-dependent problem of $\gamma$-ray transport, heating of the
ejecta gas, and transport of the radiation. Ideally, this should be
done in NLTE, and possibly in 3D to account for deviations from
spherical symmetry which are probably not rare occurrences in
SNe~Ia.

This is a complex problem, and it heavily relies on using a
predefined model of the explosion. While the development of a code
as briefly outlined above is under way, to meet also the
requirements of the 3D explosion calculations which are becoming
available from various groups, we nevertheless want to obtain as
much information as possible from the observational data sets that
are already available.

SNe have the unique feature that they unfold their content before
our eyes: as they expand, deeper and deeper layers are exposed and
contribute to line formation. Fitting a closely knit time series of
spectra offers therefore the opportunity to study the properties of
the ejecta as a function of depth, in what may be viewed as a sort
of CAT scan. Very early observations, covering the first week after
the explosion, are particularly important. At these epochs the
ejecta at velocities $\ga 18,000$\kms\ are dense enough to produce
spectral lines. Therefore it is possible to derive information about
the material at the surface of the WD. Also, it may be possible to
detect signatures of an interaction between the ejecta and
circumstellar material, which may originate from the WD or the
companion star, thus shedding light on the properties of the
progenitor system. The detail and the amount of information that can
be derived from this procedure clearly depends on the number of
spectra that are available, and on the size of the time steps.
Physical conditions in the first 8 -- 10 days after the explosion
change very rapidly compared to later epochs. Also, information
about both the progenitor and the possible interaction with
circumstellar environment is present in the outermost layers of the
ejecta. Therefore, it is desirable to have frequent very early
observations, while a lower frequency is sufficient near maximum
light and later. SN~2002bo has early and frequent data, and is
therefore very well suited for this type of study.

The best way to model the spectra independently of a particular
explosion model is to use a simple but flexible code, that can be
adapted to extract information from the spectra: we base model
calculations in the photospheric epoch on a Montecarlo (MC) code
developed in a series of papers \citep{al85,ml93,l99,m00}, and
successfully applied to various SNe~Ia \citep{m93,m95,m05a}.

The code uses a sharp lower boundary, below which all energy is
assumed to be deposited, and follows the propagation of energy
packets in a spherically symmetric envelope. The density in the
envelope depends on a model of the explosion, but abundances can be
arbitrarily chosen to fit the spectra. Ionisation and excitation
conditions are computed using a modified nebular approximation,
which was found to be a good approximation to NLTE results
\citep{pau97}. Energy packets can undergo electron scattering and
line absorption followed by reemission. The latter is treated using
the Sobolev approximation, which is appropriate for a SN envelope,
and the process of photon branching. Finally, the emergent spectrum
is computed using a formal integral approach \citep{l99}.

For the calculations in this paper, we have adopted the W7
density-velocity structure \citep{n84}. This profile is similar in
most computed models, whether deflagrations or delayed detonations.
Other inputs required by the code are the photospheric velocity
$v_{ph}$, the emergent luminosity $L$, the epoch $t_{exp}$ (time
since explosion), and the abundances.  In order to study the
abundance distribution in the ejecta, we have modified
the code to allow for radially varying abundances. The detailed
procedure is as follows.

An abundance structure is set up where the  shells are the
photospheric radii of the observed spectra. Therefore, the position
of this additional grid is not fixed a priori but is set during the
model calculations. These start with the earliest spectrum, since it
is the one with the highest photospheric velocity. A homogenous
abundance distribution above this first photosphere may be assumed,
or a number of supplementary shells (typically $\le 3$) may be
introduced at higher velocities. An optimal model is then computed
and its chemical composition, as well as the velocities of the
boundaries of the abundance shells, are stored. At the epoch of the
next spectrum in the time series the photospheric velocity has
decreased, and deeper parts of the ejecta contribute to forming the
spectrum. Although some conditions (density, temperature, etc.) have
changed in our previously considered shells, the relative abundances
have not, except for elements in the radioactive chain \Nifs\
$\rightarrow$ \Cofs\ $\rightarrow$ \Fefs, for which the intervening
decay is taken into account. Consequently, only one new abundance
shell is introduced, with boundaries given by the photospheric
velocities of the previous and the current model.  The abundances in
the new shell are set so as to achieve a best fit to the new
spectrum, while the previously determined abundances in the outer
shells are retained. This procedure is carried out until all
photospheric-phase spectra are modelled, yielding an abundance
profile of the object.

The main weaknesses of the code are the assumptions that the
photosphere is sharp and that the radiation at the photosphere has
the form of a black body. However, these assumptions are reasonable
at phases up to about maximum light, since most $\gamma$-ray
deposition occurs below the assumed photosphere, as discussed in
Section 8.

%
\section{Spectral Analysis}\label{analysis}

From the database presented in \cite{be04} we have selected for
modelling the spectra listed in Table~\ref{spec_tab}.  The two
infrared spectra already discussed in \cite{be04} are not included
in this work. In this section the analysis of each spectrum is
presented in chronological order.

\begin{table}
\caption{Spectroscopic observations of SN~2002bo used for model
calculations} \label{spec_tab}
\begin{tabular}{lccclr}
\hline \hline
Date    & M.J.D.    & Epoch$^*$ &   Range   & Tel.  & Res.\\
        &           &   (days)  &   (\AA)   &       &(\AA)\\
\hline
10/03/02& 52343.06  &  --12.9   & 3600-7700 & A1.82 & 25  \\
10/03/02& 52343.99  &  --12.0   & 3600-7700 & A1.82 & 25  \\
11/03/02& 52344.99  &  --11.0   & 3400-7700 & A1.82 & 25  \\
13/03/02& 52346.91  &  --9.1    & 3400-7700 & A1.82 & 25  \\
15/03/02& 52348.04  &  --8.0    & 3400-9050 & NOT   & 14  \\
16/03/02& 52349.93  &  --6.1    & 3400-9050 & NOT   & 14  \\
18/03/02& 52351.85  &  --4.1    & 3200-7550 & WHT   & 2   \\
19/03/02& 52352.05  &  --3.9    & 3400-9050 & NOT   & 14  \\
19/03/02& 52352.94  &  --3.1    & 3400-7700 & A1.82 & 25  \\
20/03/02& 52353.90  &  --2.1    & 3400-9050 & NOT   & 22  \\
21/03/02& 52354.96  &  --1.0    & 3400-9050 & NOT   & 22  \\
23/03/02& 52356.08  &   +0.1    & 3400-10350& A1.82 & 25  \\
28/03/03& 52361.94  &   +5.9    & 3100-8800 & WHT   & 12  \\
11/12/02& 52619.34  &   +263    & 3350-7400 & E3.6  & 14  \\
26/03/03& 52724.10  &   +368    & 3600-8600 & VLT   & 11  \\
\hline
\end{tabular}
* - relative to the estimated epoch of B maximum (MJD=52356.0) \\
A1.82 = Asiago1.82m telescope + AFOSC\\
NOT = Nordic Optical Telescope + ALFOSC\\
WHT = William Herschel Telescope + ISIS\\
E3.6 = ESO 3.6m Telescope + EFOSC2\\
VLT = ESO-VLT-U1 Telescope + FORS1\\
\end{table}

Global parameters are the distance, $\mu = 31.67$, and the
reddening.  Although \citet{be04} favour a value
$E(B-V)_{obs} = 0.43 \pm 0.10$ based on a comparison to other SNe~Ia, 
a parameter study carried out with
synthetic spectra and also presented in that paper indicates a lower
value. The highest value that is acceptable from that study is
$E(B-V)_{mod} = 0.38$.  This value lies well within the error range
of $E(B-V)_{obs}$, and is used for the calculations in this paper.
The epochs of the spectra were derived from a bolometric risetime of
$18\,$d. Table~\ref{ana_res} lists the various input parameters for
the photospheric-epoch models, and the values of the converged
temperature at the photosphere, $T_{BB}$.

\begin{table}
\caption{Input parameters and calculated converged temperature of
photospheric models.} \label{ana_res}
\begin{tabular}{ccccr}
\hline \hline
Epoch$^*$~&~Epoch$^{**}~$&~~Radius~~& ~Bol. Lum.~ &~~Temp.~\\
         & $t_{exp}$  & $v_{ph}$ & log$_{10}L$ & $T_{BB}$~ \\
   (d)   &    (d)     &  (\kms)  &   \ergs\    &    (K)~~  \\
\hline
 $-12.9$ &    5.1     &  15,800  &    42.04    &    9430   \\
 $-12.0$ &    6.0     &  15,500  &    42.31    &    9950   \\
 $-11.0$ &    7.0     &  15,100  &    42.54    &   10,930  \\
  $-9.1$ &    8.9     &  13,900  &    42.77    &   11,850  \\
  $-8.0$ &   10.1     &  12,900  &    42.84    &   12,330  \\
  $-6.1$ &   11.9     &  11,450  &    42.97    &   13,070  \\
  $-4.1$ &   13.9     &  10,400  &    43.04    &   13,310  \\
  $-3.9$ &   14.1     &  10,200  &    43.05    &   13,480  \\
  $-3.1$ &   15.0     &   9900   &    43.08    &   13,420  \\
  $-2.1$ &   15.9     &   9200   &    43.09    &   13,820  \\
  $-1.0$ &   17.0     &   8600   &    43.10    &   13,940  \\
  $+0.1$ &   18.1     &   8100   &    43.09    &   13,750  \\
  $+5.9$ &   24.0     &   7600   &    43.08    &   10,940  \\
\hline
\end{tabular}\\
$^{*}$ relative to the estimated epoch of $B$-maximum
(MJD=52356.0)\\
$^{**}$ days after the explosion.
\end{table}

\subsection{Day --12.9}\label{-12.9}

\begin{figure*}
\centering
\includegraphics[angle=270,width=.85\textwidth]{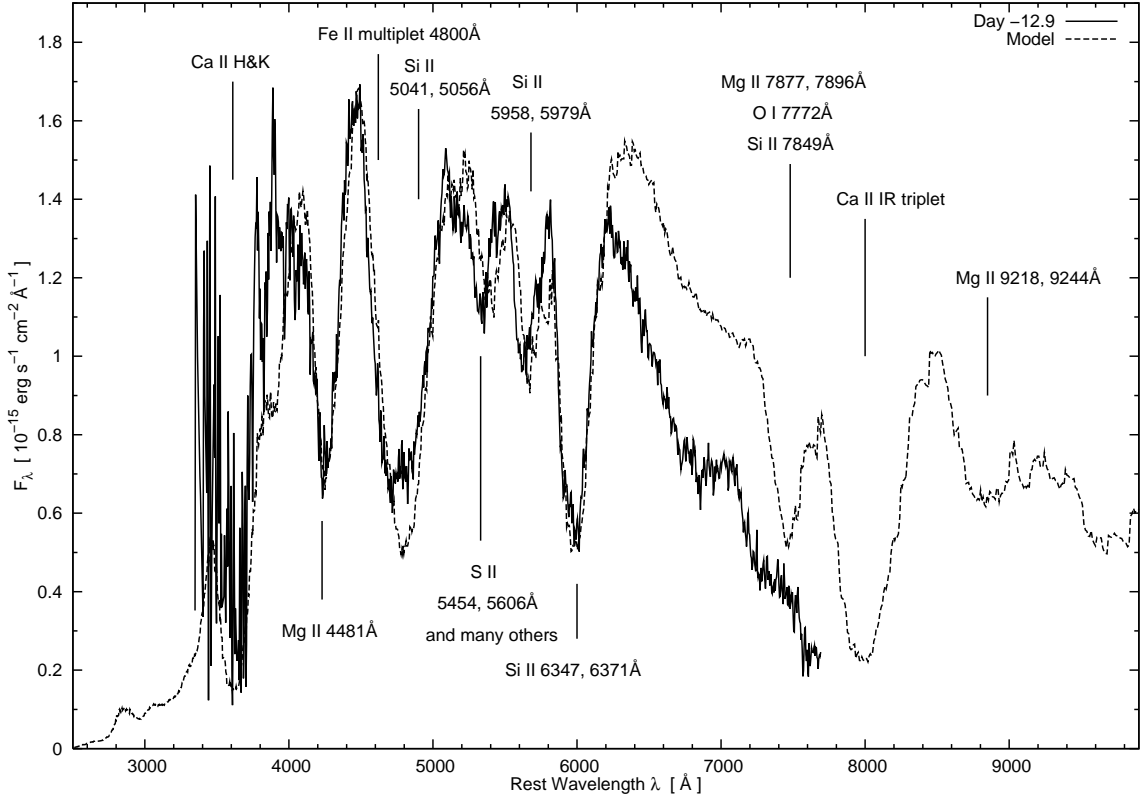}
\caption{Observed spectrum and corresponding model of SN~2002bo at
day $-12.9$ using $E(B-V)=0.38$, $\mu = 31.67$, $t_{exp} =5.1$\,d
for the model calculation. Both, observed spectrum and model are in
observer frame.} \label{min12_9}
\end{figure*}

The first spectrum in the sequence dates $12.9$ days before maximum
light, corresponding to $t_{exp} = 5.1$\,d. This is one of the
earliest SN~Ia spectra ever taken.  Fig.\,\ref{min12_9} shows the
observed spectrum and the best fit model.  The SN is still faint,
requiring $\log_{10}L=42.04$\ergs, but the photospheric velocity is
high, $v_{ph}=15,800$\kms, resulting in a low effective temperature
($T_{eff}=7540$\,K).

Several deep absorptions dominate the spectrum and are labelled in
Fig.~\ref{min12_9}. Most of these are present throughout the
photospheric phase.  The flat overall shape of the spectrum and the
ratio of the two \SiII\ features near  5700 and 6000\AA\
\citep{nu95} are indicative of a low temperature. The feature
near 7500\Ang\ (a blend of \OI\ 7772\Ang, \SiII\ 7849\Ang\ and
\MgII\ 7877, 7896\Ang), and the \CaII\ IR triplet (8498, 8542,
8662\Ang) near 8000\Ang\ are not blended in the synthetic spectrum,
but they may be in the data.

The near-photospheric composition includes a large fraction of O
(30\% by mass), but also remarkably high abundances of IME
(including 30\% Mg, 30\% Si and 6\% S), some stable Fe (1.5\%), and
1.13\% of \Nifs\ and decay products. It also includes 0.02\% of Ti
and Cr. The high Mg abundance was necessary to reproduce the feature
at $\sim 4200$\Ang. Ti and Cr are necessary to block the near-UV
flux and transfer it to redder wavelengths.  In order to reproduce
the blue extension of the \SiII\ 6355\AA\ line an additional abundance
shell was introduced above 22,700\kms. This outer shell contains
mostly Si (55\% by mass) but also significant amounts of Mg (30\%),
S (6\%) and Ca (5.5\%), so that the abundance of O is quite small.
The high Ca abundance is required to reproduce the \CaII\ lines
observed at later epochs. \2bo shows no sign of C, even at the
highest velocities.  From the present analysis, and from that of the
IR spectra \citep{be04}, we can set an upper limit for the C
abundance of 3\% restricted to velocities $>25,000$\kms. C is not
seen at later epochs either, and so we can conclude that essentially
all C (but not O) was burned during the explosion.

The main shortcoming of the model is the overestimate of the flux
beyond about 6200\Ang. Since the epoch of this spectrum is so early
that the assumption of a sharp photosphere should be correct, it is
possible that data calibration may be the reason for the mismatch of
the flux.  Spectra of other SNe~Ia at this early phase do not suffer
from this problem. Another possibility may be the adopted high
extinction. However, observations suggest $E(B-V)_{tot} > 0.33$,
while a value $E(B-V)_{tot} \la 0.25$ would be needed to improve the
situation significantly. However, such a low value would lead to
an improved model only at this particular epoch.

\subsection{Day --12.0}\label{-12.0}

\begin{figure*}
\centering
\includegraphics[width=.97\textwidth,height=.97\textheight]{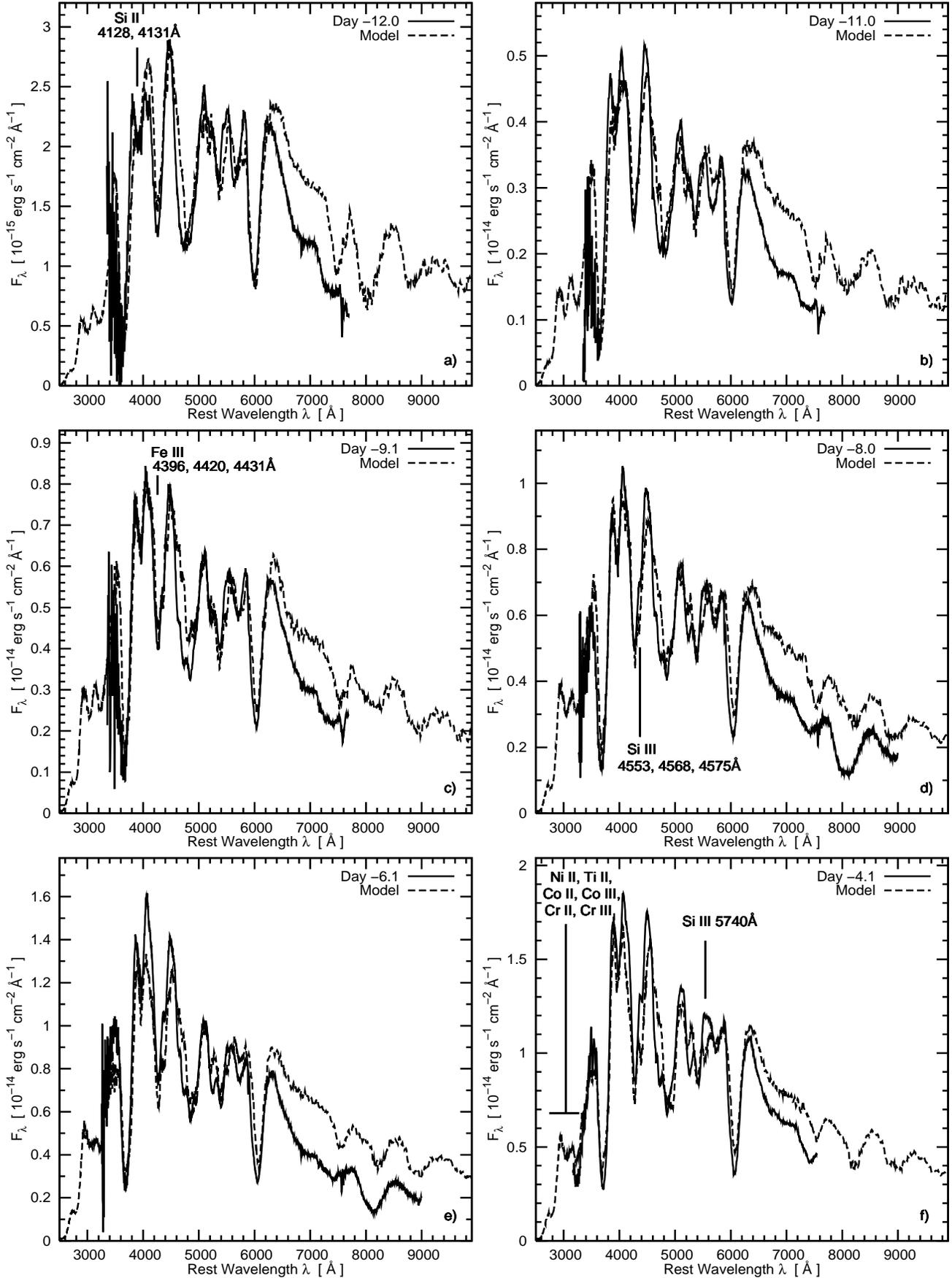}
\caption{Observed spectra and corresponding models of SN~2002bo.
Input values for the models are: $E(B-V)=0.38$, $\mu = 31.67$, a)
$t_{exp} = 6.0$\,d, b)~$t_{exp} = 7.0$\,d, c) $t_{exp} = 8.9$\,d,
d) $t_{exp} = 10.1$\,d, e) $t_{exp} = 11.9$\,d, f) $t_{exp}
=13.9$\,d. All models and observed spectra are in observer frame.}
\label{models1}
\end{figure*}

The next spectrum was taken one night after the first one at an
epoch  $t_{exp} = 6.0$\,d. The best-fit model is shown in
Fig.~\ref{models1}a. The photosphere moved inwards in velocity space
and is now located at $v_{ph} = 15,500$\kms. The luminosity is
log$_{10}L = 42.31$\ergs. Although the evolution of the spectrum is
quite small, it did become bluer: the peaks in the $U$- and
$B$-bands are higher with respect to those in the $V$- and
$I$-bands. The temperature changed only marginally, as indicated by
the \SiII\ line ratio, whereas the model temperature increased by
$\sim 500$\,K, to 8150\,K. This leads to a smaller \SiII\ line ratio.

Almost every synthetic line lacks some blue-wing absorption.  This
is best seen in the \SiII\ and \FeII\ lines. Although the Si
abundance in the outer regions is enhanced (see Sect. \ref{-12.9})
and it is very high near the photosphere (30\% by mass), this is
still not sufficient to fit the blue wings of the lines. The line
velocity required to account for the blue absorption is $\ga
20,000$\kms. To reproduce the observations may require an increase
of the density at those velocities, a situation similar to that of
most other SNe~Ia with early data \citep{m05a,m05b}.

There is still a discrepancy in the red flux between the model and
the observations, although perhaps somewhat smaller than in the
previous model.

The abundances are similar to the previous epoch, but there is a
tendency for O to decrease and for most IME (except Si and S) to
increase near the photosphere.

\subsection{Day --11.0}\label{-11.0}

The next spectrum in the series, shown in Fig.~\ref{models1}b, was
observed  one day later ($t_{exp} = 7.0$\,d). The luminosity
increased by almost a factor of 2 with respect to the previous epoch
to log$_{10}L = 42.54$\ergs. The photosphere receded to $v_{ph} =
15,100$\kms, leading to  $T_{eff}=8750$\,K. The ionisation is
consequently increased. The near-photospheric abundances of IME and
Fe-group elements are further increased in this model. Although Si
dominates near the photosphere, the synthetic \SiII\ lines
are somewhat weak. The problem is that from $v_{ph}$ up to $\sim
20,000$\kms\ more than 99.9\% of Si is doubly ionised. Thus, if the
Si abundance is further increased, \SiIII\ lines appear that are not
seen in the observations. On the other hand, the strongest \SiII\
lines are saturated near the photosphere and do not become stronger,
while outer regions do not contribute since the density is too low.
This problem will become more serious at later epochs.

\subsection{Day --9.1}\label{-9.1}

Figure~\ref{models1}c shows the spectrum at day $-9.1$ ($t_{exp} =
8.9$\,d). The photosphere was located at $13,900$\kms, and the
luminosity increased significantly, to log$_{10}L = 42.77$\ergs, so
that  $T_{eff}=9230$\,K. The \SiII\ line ratio is well reproduced,
indicating that the temperature is correctly evaluated. The weak
\SiII\ line at 4130\Ang\ is also well reproduced.

The feature near 4200\Ang, which was dominated by \MgII\ at earlier
epochs, now has a significant contribution from \FeIII\ 4420\Ang. Fe
is now doubly ionised over almost the entire envelope, which may be
an overestimate since the broad \FeII\ absorption near 4800\Ang\ is
too weak in the model.

The synthetic \CaII\ H\&K and \SiII\ 6355\Ang\ lines lack high
velocity absorption. Since the abundances at higher velocities are
constrained by the previous epochs, a possible solution is an
increased density in the outermost regions of the ejecta, as may
occur if circumstellar interaction takes place.

The abundances continue to evolve as before, with O decreasing and
Fe-group elements increasing. Among the IME, Ca and S increase, but
Mg decreases, recovering more normal values after the large increase
at the highest velocities.

\subsection{Day --8.0}\label{-8.0}

At day $-8.0$ ($t_{exp} = 10.1$\,d, see Fig.~\ref{models1}d) the
photosphere reached a velocity of $12,900$\kms. The luminosity
increased to log$_{10}L = 42.84$\ergs, and the effective temperature
is very similar to the previous epoch ($T_{eff} = 9390$\,K). The
\SiIII\ line near 4400\Ang, which was visible in the models from day
$-11$, now begins to appear in the observations. This spectrum is of
particular interest because it is the first one in the sequence that
covers the \CaII\ IR triplet. The synthetic line appears overall
somewhat weak, but the lack of blue absorption is particularly
evident. A similar behaviour can be seen in the deep absorption near
7500\Ang: the red part of this feature may be a little overestimated, 
but the blue wing is missing, probably owing to the weakness of the
high velocity Si absorption. Other \SiII\ lines confirm this effect.

The trend of the abundances is as at the previous epoch, but now
both Si and Mg decrease with O. Stable Fe is not necessary at the
velocities probed by this spectrum.

\subsection{Day --6.1}\label{-6.1}

The spectrum on day $-6.1$ ($t_{exp} = 11.9$\,d) is shown in
Fig.~\ref{models1}e. A luminosity of log$_{10}L = 42.97$\ergs, and
$v_{ph} = 11,450$\kms\ result in $T_{eff} = 9860$\,K. This is 470\,K
higher than in the previous epoch, leading to a smaller \SiII\ line
ratio and to a stronger \SiIII\ line near 4400\Ang. The \SII\
absorption near 5300\Ang\ is weaker in the model than in the
observations. This is a known problem from previous analyses of
SNe~Ia, possibly caused by uncertain $gf$-values.

The abundance of O and Si is unchanged, those of \Nifs, S and Ca
increase, compensated by the decrease of Mg.

\subsection{Day --4.1}\label{-4.1}

The spectrum at day $-4.1$ (Fig.~\ref{models1}f) has $t_{exp} =
13.9$\,d.  The model was computed with log$_{10}L = 43.04$\ergs,
$v_{ph} = 10,400$\kms, and it has $T_{eff} = 10,040$\,K.

The model matches the observation very well in the blue, showing the
ability of the MC code to treat line blanketing. The absorption at
the blue edge of the spectrum is a blend of many \NiII, \CoII,
\CoIII, \TiII, \CrII\ and \CrIII\ lines, which cannot be
distinguished unambiguously.  Differences between the model and the
observations are due to rather strong lines of \SiIII\ $(\sim
4400$\Ang\ and 5500\Ang), suggesting that the temperature may be somewhat 
too high, and to missing high-velocity absorption ($v \ga 18,000$\kms), most 
prominently in \SiII\
6355\Ang\ and in the \MgII-dominated feature near 7400\Ang. The
problem with \SII\ was discussed in Section~\ref{-6.1}.

Interestingly, the red flux discrepancy is much smaller in this WHT
spectrum than in all previous ones, which were from Asiago and NOT.
Maybe the assumption of a photospheric black body is more realistic
here, but the observed spectrum is also significantly different from all
previous ones. Therefore, questions about the calibration remain.

The abundance of O and Mg is further decreased, but other IME remain
constant, having reached the maximum of their abundances. \Nifs\
increases further, and is the dominant element near the photosphere.

\subsection{Day --3.9}\label{-3.9}

\begin{figure*}
\centering
\includegraphics[width=.97\textwidth,height=.97\textheight]{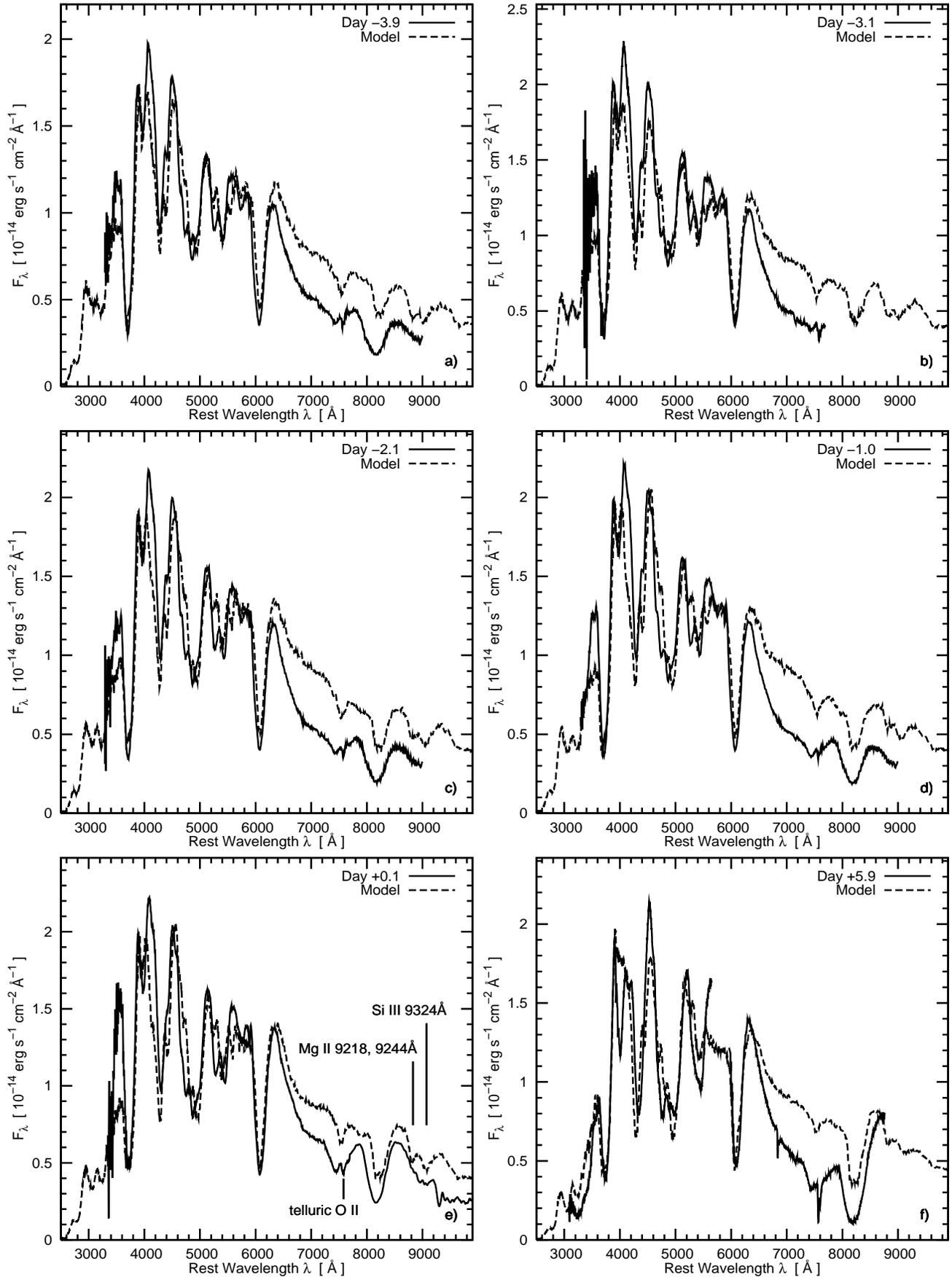}
 \caption{Observed spectra and corresponding models of SN~2002bo.
  Input values for the models are: $E(B-V)=0.38$, $\mu = 31.67$,
  a) $t_{exp} = 14.1$\,d, b) $t_{exp} = 15.0$\,d, c) $t_{exp} = 15.9$\,d,
  d) $t_{exp} = 17.0$\,d, e) $t_{exp} = 18.1$\,d, f) $t_{exp} = 24.0$\,d.
  All observed and modelled spectra are in observer frame.}
 \label{models2}
\end{figure*}

The spectrum at day $-3.9$ is shown in Fig.~\ref{models2}a. The
model has $t_{exp} = 14.1$\,d, $v_{ph} = 10,200$\kms, log$_{10}L =
43.05$\ergs\ and $T_{eff} = 10,080$\,K.

Although this spectrum was observed only 4.8\,h after the one on day
$-4.1$, it is significantly bluer. Spectral evolution is unlikely to
be so rapid, highlighting the problems of calibration. The two
spectra were observed with different telescopes. The exact colour
evolution $\Delta (B-V)$ is not known because no photometry is
available for the epoch of the day $-3.9$ WHT spectrum, and only a
$B$-band measurement exists at the epoch of the NOT spectrum. The near
photospheric composition is insignificantly different from the
previous epoch. The model fails to reproduce the bright blue peaks.
This may be due to the black body distribution placing too much flux
in the red (Sect.~\ref{-6.1}), but relative flux calibration may
again be questionable.

\subsection{Day --3.1}\label{-3.1}

The spectrum at day $-3.1$ ($t_{exp} = 15.0$\,d) is shown in
Fig.~\ref{models2}b. The luminosity log$_{10}L = 43.08$\ergs\
increases more slowly compared to earlier epochs, as the light curve
approaches its peak. The velocity decrease is also slower ($v_{ph} =
9900$\kms). Consequently, the temperature at the photosphere is the
same as in the previous epoch ($T_{eff} = 10,090$\,K). The small \SiII\
line ratio at this epoch is reproduced very well by the model.

Because of the small step in $v_{ph}$, the abundances are practically
unchanged near the photosphere, except for a small increase of
\Nifs\ and a decrease of O, Si and Ca. A high Si abundance is still
necessary to fit the \SiII\ 6355\Ang\ absorption.

\subsection{Day --2.1}\label{-2.1}

The spectrum at day $-2.1$, shown in Fig.~\ref{models2}c, was taken
at an epoch of $t_{exp} = 15.9$\,d. The luminosity (log$_{10}L =
43.09$\ergs) increased only slightly with respect to the previous
day. The photosphere receded to $v_{ph} = 9200$\kms, leading to a
somewhat higher temperature, $T_{eff} = 10,210$\,K. The relative
depth of the \MgII\ line near 7500\Ang\ is now correctly reproduced,
but both this line and \CaII\ IR lack blue absorption.

The O abundance is further decreased, compensated by an increase of
\Nifs. Some stable Fe is again required, and all IME decrease.

\subsection{Day --1.0}\label{-1.0}

The last spectrum before maximum light has a fiducial epoch of
$t_{exp} = 17.0$\,d (Fig.~\ref{models2}d). The photospheric velocity
$v_{ph}$ receded to 8600\kms. The bolometric luminosity reached its
maximum, log$_{10}L = 43.10$\ergs, at this epoch, one day before $B$
maximum. The temperature also reached its maximum, at a value
$T_{eff} = 10,250$\,K. The day $-2.1$ and the day $-1.0$ spectra
look very similar, but the latter is slightly bluer. Unfortunately,
only $B$-band photometry is available to calibrate the spectra. A
blend of \CoIII, \FeIII\ and some weaker \SIII\ lines produces the
deep absorption feature near 4300\Ang. The abundances of these
elements may be slightly too high but cannot be reduced without a
negative influence on the overall shape of the synthetic spectrum. Aundances 
have the same trend as in the previous epoch.

\subsection{Day +0.1}\label{0.1}

The maximum light spectrum is shown in Fig.~\ref{models2}e. An epoch
of 18.1 days is assumed. The photospheric velocity decreased to
8100\kms, and the luminosity already begun to decline at log$_{10} L
= 43.09$\ergs. The expansion and the declining luminosity reduce the
temperature to $T_{eff} = 10,200$\,K.

It is again interesting that this spectrum, the only one extending
beyond 9000\Ang, is also the one whose red flux is best reproduced
by the model, even though the photospheric approximation is
certainly less valid at maximum than at earlier epochs. This looks
like rather convincing evidence that most spectra that do not extend
sufficiently to the red are affected by significant red-end
calibration problems.

The depth of the \CaII\ IR triplet relative to the continuum is
reproduced very well, and only the blue wing of the line is missing.
\SiII\ lines are also accurately reproduced, but there are unwanted
\SiIII\ lines at various places in the spectrum (e.g. 4400,
5600\AA). The observed spectrum exhibits a much stronger peak $\sim
3500$\Ang\ than the previous one. The somewhat erratic relative
behaviour of the blue and the red part of the spectrum again
suggests that there may be inconsistencies in the data calibration.

Abundances change smoothly, following the trend of decreasing O and
IME, increasing \Nifs\ and stable Fe.

\subsection{Day +5.9}\label{5.9}

The last spectrum is from day +5.9, corresponding to $t_{exp} =
24.0$\,d. The observed spectrum consists of two distinct parts, as
shown in Fig.~\ref{models2}f, making it difficult to tell whether
the high emission near 5500\Ang\ is real and whether the relative
flux level of the two parts is correct.  The blue part ends just to
the red of the \SII\ feature, which appears to be blended.

Although the epoch is several days after maximum, the luminosity has
decreased only slightly (log$_{10}L = 43.08$\ergs). This may be a
calibration problem. The model velocity decreased to $v_{ph} =
7600$\kms, so the temperature dropped significantly ($T_{eff} =
9100$\,K) as a consequence of the expansion.  The synthetic spectrum
reproduces the overall flux distribution and almost all of the
lines.

This spectrum explores sufficiently deep layers that the derived
abundance distribution overlaps with that from the models of the
nebular phase (Section~\ref{neb}). However, the assumption of a
sharp photosphere is weakest at this epoch, making the derived
abundances more uncertain. Nevertheless, the fact that the
abundances continue to change in the same way as at earlier epochs
is comforting.

%
\section{Nebular phase}\label{neb}
\begin{figure*}
\centering
\includegraphics[width=\textwidth]{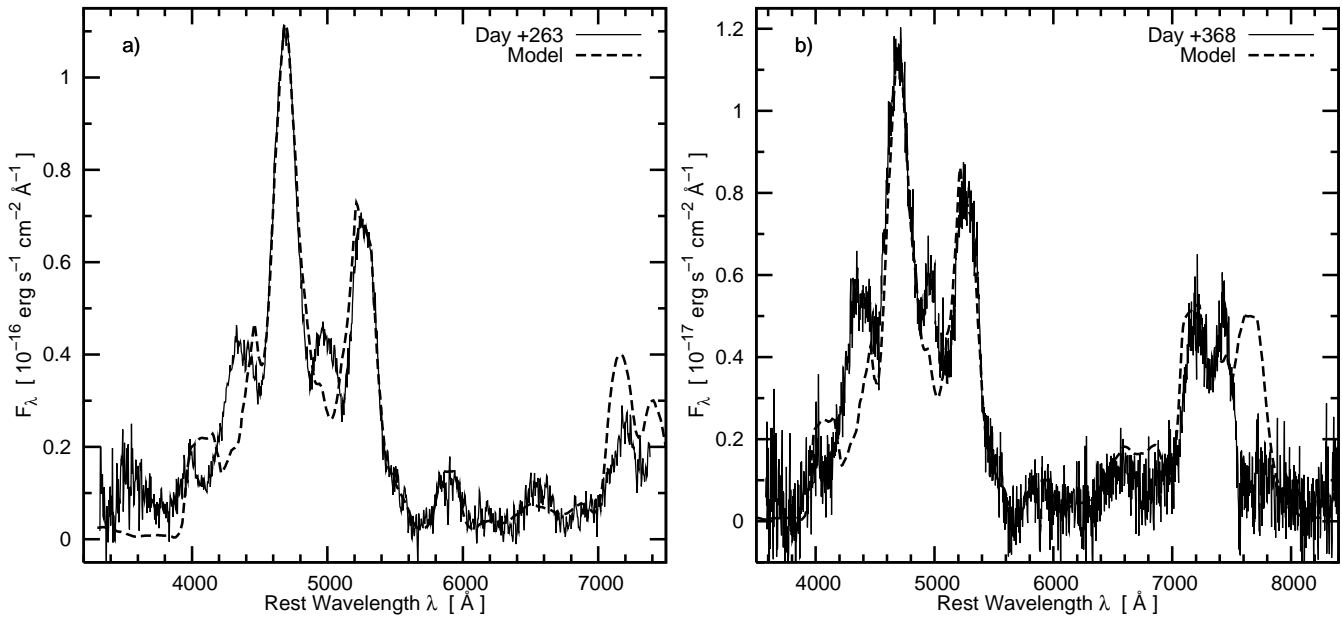}
 \caption{Observed spectra and corresponding models of \2bo\ in the
 nebular phase. Input values are: $E(B-V) = 0.38$, $\mu = 31.67$,
a) $t_{exp} = 282$\,d, b) $t_{exp} = 386$\,d.} \label{nebspec}
\end{figure*}

The photospheric epoch coverage of \2bo\ ends at day $+6$, when
$v_{ph} = 7600$\kms. It is therefore not possible to extend the
analysis to lower velocities using the MC code. In any case, the
code becomes less and less applicable at advanced post-maximum
epochs, as the photospheric approximation becomes rapidly
inadequate. There is, however, an alternative approach that makes
it possible to investigate the properties of the inner part of the
ejecta. That is modelling the spectra in the so-called nebular
phase, when the ejecta are completely transparent. Models
applicable to this epoch have been developed, and applied to
various types of SNe \citep[e.g.][]{m01_2}. These models follow
the energy deposition and heating by the $\gamma$-rays and the
positrons from the radioactive decay of \Nifs\ and \Cofs, and the
ensuing cooling by line emission.

We extended our previous one-zone model also to treat abundance
stratification. The $\gamma$-ray deposition is treated in a
Montecarlo approach \citep[see][]{ca97}, while nebular emission is
computed in NLTE. This code can be used to reproduce accurately the
emission line profiles, and hence to derive the abundance
distribution in the ejecta.

Two nebular spectra are available, taken on Dec. 11th, 2002 and
March 26th, 2003, i.e. 282 and 386 days after the estimated
explosion epoch, respectively (see Tab.~\ref{spec_tab}). They were
reduced with standard IRAF\footnote{The Image Reduction and
Analysis Facility (IRAF) is distributed by the National Optical
Astronomy Observatory, which is operated by AURA, Inc., under a
cooperative agreement with the National Science Foundation.}
routines. Extractions were weighted by the variance based on the
data values and a Poisson/CCD model, using the gain and read noise
parameters. The background to either side of the SN signal was
fitted with a low-order polynomial and then subtracted. Flux
calibration was performed using spectrophotometric standard stars.

We modelled the spectra adopting the W7 density profile, fixing the
abundances in the outer part of the ejecta according to the results
of the photospheric epoch study above. Thus we only had the freedom
to modify abundances below 7600\kms.

The model for the day $+263$ spectrum is shown in
Fig.~\ref{nebspec}a. It reproduces the observed spectrum very well
if the abundance of \Nifs\ at velocities below 7600\kms\ is
increased above the values estimated from the photospheric epoch
study at higher velocities. However, the \Nifs\ abundance reaches at
most 0.62 by mass in regions between 4000 and 7000\kms. This
confirms that significant mixing out of \Nifs\ occurred, as derived
from the early-time spectra. Further inwards, the \Nifs\ abundance
decreases, as expected. If the innermost ejecta contain mostly
\Nifs, the Fe-line emission becomes too strong. However, line
profiles show that there is Fe emission at the lowest velocities.
This can be reproduced if the innermost layers contain significant
amounts of stable Fe ($\sim 0.30$\,\M\ at $ v < 5000$\kms). Stable
Fe must be excited by (mostly) $\gamma$-rays and (almost negligibly)
positrons emitted at higher velocity layers and diffusing inwards.
Stable Fe should be mostly \Feff\ and \Fefe. These isotopes are the
result of burning to NSE, but at slightly different densities.
Spectroscopically, we are unable to distinguish between them. A
total \Nifs\ mass of $0.50$\,\M\ was used, which makes \2bo\ an
average SN~Ia. The \Nife\ abundance is low.  A total $\sim
0.80$\,\M\ was burned to NSE. The abundance distribution is plotted
in Fig.5.

Synthetic spectra for the later epoch (day $+368$) computed using
the same abundance distribution nicely confirm these results
(Fig.~\ref{nebspec}b).

Significant changes of the line profiles or width over time are not
visible in the data.

%
\section{Abundance Stratification}\label{at}

\begin{figure*}
\centering
\includegraphics[width=.95\textwidth,height=.97\textheight]{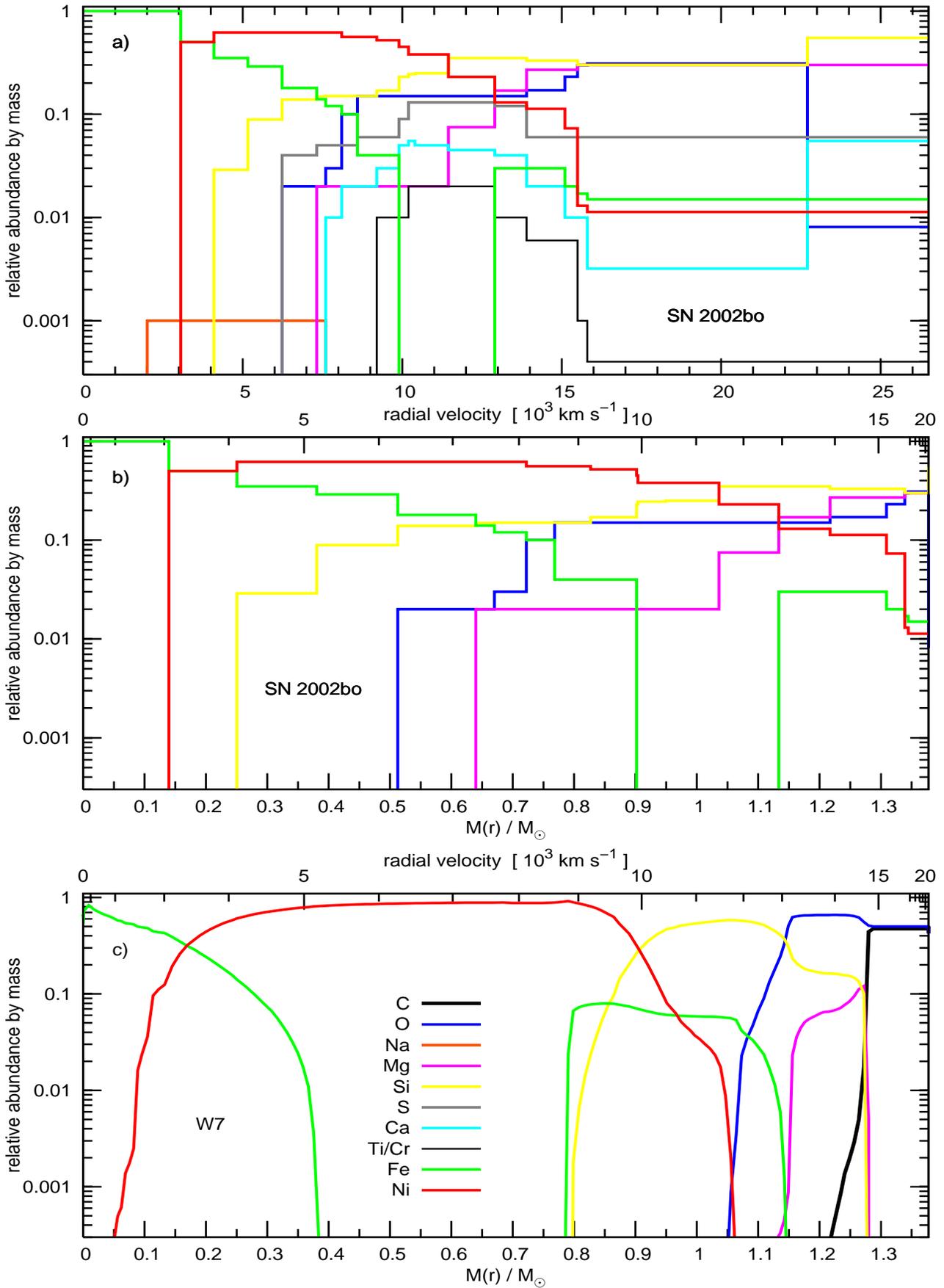}
 \caption{{\em Upper panel:} Abundance distribution of \2bo including
  all elements. {\em Middle:} O, Mg, Si, Fe, and \Nifs\ abundances in
  \2bo. {\em Lower panel:} W7 deflagration model abundances (C, O, Mg, Si,
  Fe, \Nifs).}
 \label{abun}
\end{figure*}

Modelling of all spectra, both of the photospheric and the nebular
phase, delivers the abundance distribution of the entire ejecta. The
relative abundance by mass of the elements that could be detected in
the spectra is shown in Fig.~\ref{abun}a as a function of radius in
velocity space.

The various steps above 7600\kms\ reflect the radii of the
early-time spectra. Therefore, the distance between two abundance
shells depends on the time interval between the two spectra. An
additional shell above $22,700$\kms\ was inserted in order to
account for high velocity material (Sect.4.1). Below 7600\kms,
where the abundances are derived from the nebular spectra, the
radii of the shells are given by the density shells of the
underlying W7 model. Since our models cover the entire SN ejecta,
the total mass of each element can be calculated. These values are
listed in Table~\ref{abun_tab}.  Different isotopes cannot be
distinguished spectroscopically, so here the abundance of an
element implies the sum of all isotopes, except for Fe.

\begin{table}
\caption{Total mass of elements spectroscopically confirmed in the
ejecta of SN~2002bo and nucleosynthesis products of different
SN~Ia explosion models} \label{abun_tab}
\begin{tabular}{ccccc}
\hline \hline
        & \multicolumn{4}{c}{Ejected (Synthesized) Mass (\M)} \\
          \cline{2-5}\\
Species & SN 2002bo & W7$^a$    & WDD1$^b$  & b30\_3d\_768$^c$ \\
\hline
    C    &$\la 0.002$& 4.83E-02  & 5.42E-03  & 2.78E-01  \\
    O    &    0.110  & 1.43E-01  & 8.82E-02  & 3.39E-01  \\
    Na   &    0.001  & 6.32E-05  & 8.77E-05  & 8.65E-04  \\
    Mg   &    0.080  & 8.58E-03  & 7.69E-03  & 8.22E-03  \\
    Si   &    0.220  & 1.57E-01  & 2.74E-01  & 5.53E-02  \\
    S    &    0.067  & 8.70E-02  & 1.63E-01  & 2.74E-02  \\
    Ca   &    0 020  & 1.19E-02  & 3.10E-02  & 3.61E-03  \\
    Ti   &    0.003  & 3.43E-04  & 1.13E-03  & 8.98E-05  \\
    Cr   &    0.003  & 8.48E-03  & 2.05E-02  & 3.19E-03  \\
  Fe$^*$ &    0.360  & 1.63E-01  & 1.08E-01  & 1.13E-01  \\
\Nifs$^{**}$& 0.520  & 5.86E-01  & 5.64E-01  & 4.18E-01  \\
Ni$^{***}$&    n.a.  & 1.26E-01  & 3.82E-02  & 1.06E-01  \\
\hline
\end{tabular}\\
$^*$Stable isotopes except for \Fefs\ from \Cofs\ decay \\
$^{**}$Isotopes from the \Nifs\ $\rightarrow$ \Cofs\ $\rightarrow$
\Fefs\ decay chain \\
$^{***}$Stable isotopes $^{58}$Ni, $^{60}$Ni, $^{61}$Ni,
$^{62}$Ni, and $^{64}$Ni\\
$^{a}\,$\citet{n84}\\
$^{b}\,$\citet{i99}\\
$^{c}\,$\citet{t04}
\end{table}

Since the progenitor WD is supposed to be composed of C and O, the
abundances of these two elements in the SN ejecta indirectly
indicate the extent of burning. Interestingly, no sign of C can be
seen in the spectra of \2bo. The most prominent C lines in the
optical and NIR wavelength bands are \CII\ 6579\Ang\ and \CII\
7231\Ang. Neither line can be detected, at any epoch.  \cite{be04}
suggested an upper limit of the C abundance of 3\% at $v >
25,000$\kms\ based on models of the optical spectra. This upper
limit corresponds to a total mass of $\la 2\times 10^{-3}$\,\M.
Since no C lines are detected in the spectra, the C abundance in
this analysis is set to zero.

The other progenitor element, O, dominates the ejecta between
$15,500$\kms\ and $22,700$\kms. In the outermost layers the
abundance of Si is higher than that of O. In deeper layers the O
abundance decreases, and no O can be found below 6000\kms.
Altogether we detect $\sim 0.11$\,\M\ of O in \2bo.

The group of IME includes Na, Mg, Si, S and Ca. Except for Na, they
all show rather high abundances at high velocities. Na itself is
located above the Fe core and out to $\sim 8500$\kms, with a low
abundance.

Mg is enhanced in \2bo. A total mass of $\sim 0.08$\,\M\ is derived
from the modelling. The bulk of Mg is at $v > 12,000$\kms, with
relative mass abundances between 20\% and 30\%. This is necessary to
fit the deep 4481\Ang\ absorption in the early phase, as discussed
in Section~\ref{analysis}. Below $\sim 15,000$\kms\ the Mg abundance
decreases steadily, and no Mg is detected below 7300\kms.

The Si abundance has a similar behaviour as Mg. However, the
increase of the abundance at high velocity is more pronounced. The
outermost shell ($v > 22,700$\kms) contains 52\% of Si in order to
account for the blue wing absorption of the various \SiII\ features.
In the region between $11,450$\kms\ and $22,700$\kms\ Si is as abundant
as O ($\sim 30$\%). This is similar to the prediction of a model
like W7, although somewhat on the high side. Deeper in the ejecta
the Si abundance is $\sim 15$\% down to $\sim 6000$\kms, and then it
drops rapidly until it disappears below 4000\kms. A total mass of
$\sim 0.22$\,\M\ is derived, which is more than in W7 but less than
in the DD models. It must be emphasised that the Si abundance in the
deepest layers can only be derived indirectly, since the only
nebular lines of Si are in the IR, which is not available for
SN~2002bo.

S behaves like Si, but with a lower abundance, and does not exhibit
a rise in the outermost shell. It starts at 6\% outside, increasing
to $\sim 13$\% between 12,900 and 10,200\kms, and decreasing to
$\sim 6$\% before disappearing below $\sim 6000$\kms. The total mass
of S is $\sim 0.067$\,\M.

Ca completes the set of IME that can be detected in \2bo. This
element is the best to study high velocity components, since its
lines are the strongest. As discussed in Sect.~\ref{analysis}, the
\CaII\ IR triplet lacks significant absorption in its blue wing. We
tried to account for this using up to 20\% Ca at intermediate and
especially at high velocities. However, even with 100\% Ca in the
outermost layers it is impossible to fit the blue part of the line
in the day --8 spectrum.  On the other hand, the strong \CaII\ 7291,
7324\Ang\ nebular emission can be reproduced with a very small Ca
abundance ($\sim 10^{-5}$ by mass). Therefore, a second iteration of
the modelling of the early time spectra was performed using a
significantly lower value for the Ca abundance. A satisfactory
result was obtained for an abundance of 5.5\% above $22,700$\kms,
decreasing to 0.3\% in the next deeper shell and increasing again to
1\% -- 5.5\% between $15,800$ and 7500\kms.  Below this velocity,
the abundance is very small. A total 0.021\,\M\ of Ca is estimated.
The problem with the high velocity absorptions cannot be solved by a
simple enhancement of the abundances, and may require scenarios like
an increased density in the outer ejecta, possibly due to
circumstellar interaction.

For Ti and Cr, which have a similar abundance trend, a total mass of
$\sim 6\times 10^{-3}$\,\M\ is derived. Although no individual lines
of these species can be detected in the spectra, their line-blocking
effect is necessary to shift UV and blue photons to longer
wavelengths. The distribution of these elements peaks between
$10,000$ and $15,000$\kms, with values between 1 and 2\%.

The abundance of \Nifs\ includes 
plus both the \Cofs\ and \Fefs\ that form in the decay chain. The
relative abundances of these species change with time. \Nifs\
dominates between about 3,000 and 10000\kms, but its abundance
never exceeds $\sim 60$\%. Beyond this velocity the abundance
decreases, dropping to $\la 1$\% above $15,500$\kms.  We
estimate the total \Nifs\ mass synthesised to be $\sim 0.52$\,\M.

The abundance of just Ni includes all stable Ni nuclei (\Nife,
$^{60}$Ni, $^{61}$Ni, $^{62}$Ni, and $^{64}$Ni), which are mostly
located at the lowest velocities. The dominant isotope should be \Nife.
The production of these species depends on the neutron excess
during the burning regime. Especially in the W7 model stable Ni
makes up a significant mass fraction of the inner ejecta. 
The synthetic spectra cannot differentiate between different isotopes.

The heading ``Stable Fe" stands for the sum of all Fe isotopes
(\Feff, \Fefs, $^{57}$Fe, \Fefe), except for the \Fefs\ that is
produced in the decay of \Cofs. A stable Fe core extends out to
$\sim 3000$\kms. Further out stable Fe decreases steadily, and it
is absent in the velocity range from 9900 to $12,900$\kms. This
means that the \Fefs\ from the decay of \Nifs\ is sufficient to
reproduce the spectra. Above this void region a small amount of
stable Fe $(\sim 2$\%) is again detected, extending to the highest
velocities. The presence of stable Fe at high velocities is
required because the outer layers are observed so early that only
very little \Fefs\ has been produced from \Nifs. Altogether a
total mass of 0.36\,\M\ of stable Fe is measured.

We compared the derived abundance distribution of SN~2002bo with
various explosion models. Figure~\ref{abun}b shows the distribution
of the main elements (O, Mg, Si, stable Fe isotopes, and \Nifs)
versus $M(r)$. The 1-D deflagration model W7 \citep{n84} is shown in
Fig.~\ref{abun}c for comparison.  The radial distribution is very
similar in the two plots. In both cases, the innermost part of the
envelope is dominated by stable Fe isotopes. Above this is the
\Nifs-dominated region, and IME are located further out.
Despite these similarities there are significant differences. In
\2bo the abundance pattern is shifted to higher velocities and the
regions of Fe~group elements, IME, and unburned material are not
as well separated but rather more mixed. The inner stable Fe
region is much larger. O is mixed down, Si and Mg extend to both
lower and higher velocities than in W7 while \Nifs\ and Fe are
mixed to the outer regions. Finally, the region above $\sim
1.3$\,\M, which is completely unburned in W7, in \2bo shows no
trace of C and both IME and Fe-group elements are mixed outwards.
In practice, in \2bo we detect more stable Fe but no stable Ni (mostly
\Nife\ in W7).

Since delayed detonation models burn high velocity regions to IME,
they could possibly compare better to the abundance distribution
of \2bo. Hence we examined various detonation models, namely WDD1,
WDD2, WDD3, CDD1, CDD2, and CDD3 \citep{i99}. Indeed, the
occurrence of IME at high velocities in \2bo is represented better
by the DD models than by W7. However, O and the Fe-group elements
show a very different behaviour, much closer to W7. In particular,
the relative locations of Fe, Ni, IME, and unburned O is
reproduced best by W7.

Therefore we may conclude that in the case of \2bo W7 is
the explosion model that fits best the abundance distribution
derived from spectral synthesis calculations. However, if mixing
effects are neglected it is difficult to reproduce the observed
abundance distribution exactly. On the other hand, the analysis of
a single object is not sufficient to constrain different explosion
models. Abundance stratification analysis of several objects is
necessary, and it is under way.


\section{Bolometric light curve}\label{lc}

\begin{figure}
\includegraphics[width=.48\textwidth]{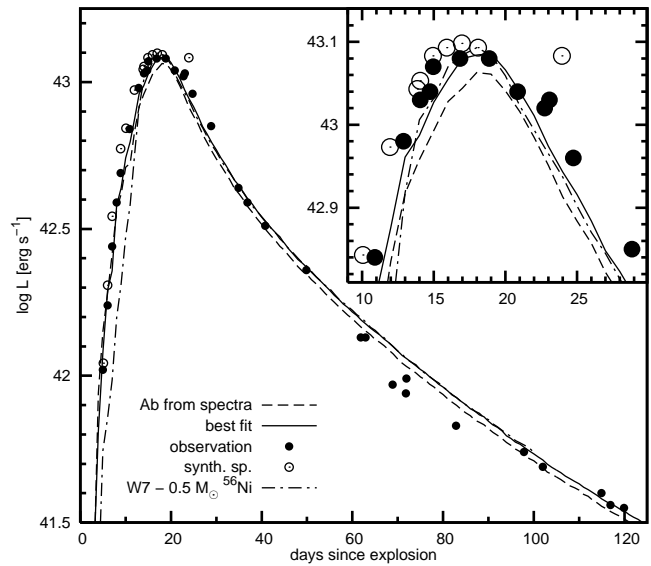}
 \caption{uvoir light curve of \2bo. The black dots represent the
 observed photometric points, the circles are the bolometric luminosity
 deduced from the synthetic spectra. Three models of the MC light curve code
 are shown using directly the abundances from the spectral analysis
 ($--$), the same abundances but with slightly increased \Nifs\ mass
 (\textemdash), and W7 abundances with 0.5\,\M\ of \Nifs\ ($-\cdot-$).}
 \label{lcfit}
\end{figure}

In order to check the validity of our results independently, we used
the abundance distribution obtained from the spectral modelling and
computed a synthetic bolometric light curve, assuming that the
density is given by W7 and using a grey MC light curve model
\citep{ca97,m01_1}. In Fig.~\ref{lcfit} we compare the results with
the bolometric light curve of SN~2002bo presented in \citet{be04}.

The synthetic light curve reproduces the observed one successfully,
and it performs exceedingly well  in the early, rising branch. This
is particularly exciting. In fact, the early light curve rise
depends mostly on the mixing out of \Nifs. A synthetic light curve computed 
with the unmixed W7 abundances and scaled to a \Nifs\ mass of
0.5\,\M\ shows a significantly later rise, which is in disagreement
with the observations (Fig.~\ref{lcfit}). In that case, \Nifs\
extends only up to $\sim 20,000$\,\kms, so photons are trapped
longer in the ejecta and the rise of the light curve is delayed.
Evidently, abundance stratification is accurately determined by the
spectral modelling, especially in the outer layers.

The model is less successful near the peak of the light curve. A
slight modification of the abundances, with \Nifs\ increased to
about 0.56\,\M\ and stable Fe decreased accordingly, yields a better
fit of this phase. This is a small change, which does not affect the
essence of our results.


\section{The position of the photosphere and the reliability of the results}
\label{vel}

\begin{figure}
\includegraphics[width=.48\textwidth]{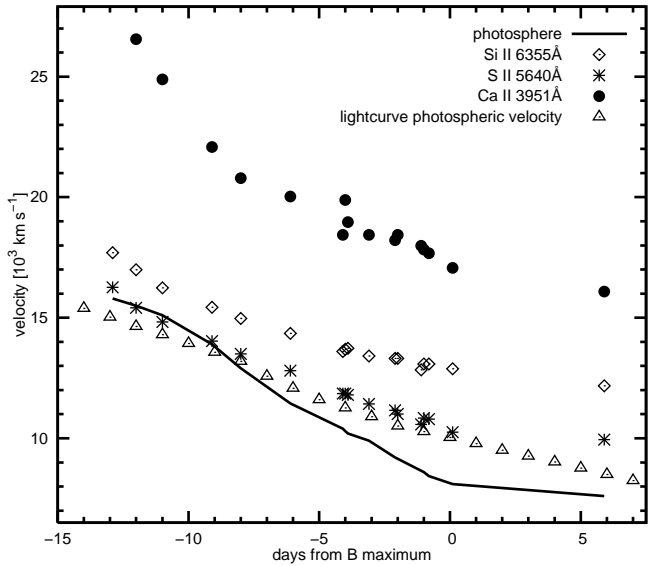}
 \caption{Evolution of the expansion velocities deduced from the
  minima of the \SiII\ 6355\Ang, \SII\ 5640\Ang, \CaII\ H\&K
  absorptions and the photospheric radius of the synthetic spectra, and the
  position of the photosphere in the light curve code.}
 \label{lvel}
\end{figure}

We mentioned that the velocities of IME lines in \2bo\ are
peculiarly high. Therefore, it is interesting to see how the
velocities of the observed lines compare to the model calculations.
Especially for the weaker lines, the photospheric radius should be a
good estimate for the expansion velocities of the model line
features. The evolution of the photospheric velocity is compared in
Fig.~\ref{lvel} to the expansion velocities of \SiII\ 6355\Ang,
\SII\ 5640\Ang, and \CaII\ H\&K, as measured from their minima
\citep{be04}.

The photosphere of \2bo\ is at 15,800\kms\ on day $-12.9$, and
recedes only by 700\kms\ until day $-11.0$. This slow decline phase
accounts for trying to reproduce the high-velocity absorption
features. After day $-11.0$, the decline is steeper (590\kmsd), and
then it slows down again after maximum. The three lines exhibit a
slightly different behaviour than the photosphere. Their velocities
decrease more rapidly until day $\sim -8$ then they do afterwards.
After maximum, the gradient is very similar to the photospheric one.

The \SII\ line has the lowest velocity, as may be expected since
this is the weakest line that can be individually measured. After
day $-8.0$ it forms well above the Montecarlo photosphere, but it matches the
light curve photosphere up to abouty maximum. 

The strong \SiII\ line declines at a slower rate, especially between
day $-6.1$ and maximum. It is always significantly faster than the
photospheric velocity, as can be expected since this strong line is
formed well above the photosphere.

The \CaII\ H\&K line is very strong and it has the highest
velocities. It evolves like the \SiII\ line, at higher velocities.
However, the very high velocities measured in the earliest spectra
($\sim 25,000$\kms) in both \CaII\ H\&K and \CaII\ IR are probably
not photospheric, but indicate the presence of a high-velocity
component \citep{m05a,m05b}.

The calculation of the light curve offers the opportunity to
investigate the validity of the assumption of a sharp photosphere
that is made in the MC code. The light curve code
in fact computes $\gamma$-ray deposition as a function of depth and
time, and estimates the position of a gray photosphere based on a
simple opacity prescription. The photospheric velocity thus
computed, also shown in Fig.6, follows closely the position
determined by the MC spectrum synthesis code, and, even more
closely, that traced by the velocity of the \SII\ line, which is
sufficiently weak to be a good tracer of $v_{ph}$. We can now look
at the fraction of the energy deposited above the photosphere as a
function of time. This fraction reaches 10\% between 1 and 2 days
before maximum, it is $\sim 13$\% at maximum and $\sim 25$\% 6 days
after maximum. These values indicate that up to about maximum the
approximation of a sharp photosphere is acceptable, while in the
spectrum at day $+5.9$ it is rather poor, and non-thermal excitation
and ionisation effects that are not considered in the code may be
relevant. Another test is the fraction of \Nifs\ located above the
photosphere. This is only $\sim 10$\% at maximum and $\sim 20$\% 6
days later, confirming the above result.

Fortunately, the depth of the zone explored between maximum light
and the nebular spectra is quite small (between 7600 and 9000\kms),
since the photosphere recedes only slowly after maximum and then
disappears. Nevertheless, abundances in that velocity range must be
regarded as more uncertain than values both at higher velocities,
where the MC code is more reliable, and at lower ones, which are
treated with the nebular code.


\section{Conclusions}\label{concl}

We have derived the abundance stratification in the ejecta of the
SN~Ia 2002bo through a time series of spectral models. Synthetic
spectra were computed for 13 epochs during the photospheric phase
and 2 in the nebular phase.

The most important result is that although the abundances are not
very different from those of a standard model such as W7, some
mixing in abundance seems to have occurred. In particular, \Nifs\
extends to higher velocities than in W7, and the IME are mixed
both inwards and outwards. Some IME are present at the highest
velocities, which may indicate interaction of the ejecta with a
CSM or some asphericity in the explosion. Balancing this, O is
mixed inwards. C is however not seen in the ejecta.

One possibility to produce a situation like that observed in \2bo\
is multi-dimensional effects. Bubbles of burned material may rise to
outer regions while fingers of unburned material sink to deeper
layers. Unlike the case of global mixing, the direction from which
the observer looks at the SN may be important. In the case of \2bo\
one of these partially burned bubbles reaching out to high
velocities may be in the line of sight, delivering the spectra and
abundance distribution we observe. Had we looked from another
position we might possibly have seen the missing C and less or no
high-velocity components in the lines of IME \citep[see,
e.g.][]{ka04}. In this context it would be interesting to verify
whether multi-dimensional effects can account for the observed
diversity in SNe~Ia beyond the brightness-decline rate relation.

The investigation of the abundance stratification in a number of
well-observed SNe~Ia can contribute to our understanding of all
these mechanisms.

\bigskip
%

\noindent {\bf AKNOWLEDGEMENTS}\\

This work is supported in part by the European Union's Human
Potential Programme under contract HPRN-CT-2002-00303, `The
Physics of Type Ia Supernovae.'
We would like to thank the Institute for Nuclear Theory at the
University of Washington, Seattle, USA, for supporting
a visit in the summer of 2004.

%
%

%
\label{lastpage}
\end{document}